\documentclass[twocolumn,floats,aps]{revtex4-1}
\usepackage{times,amsmath,psfrag,latexsym,pstricks,graphics}

\begin{document}
\title{Vortex pinning by surface geometry in superfluid helium}
\author{I.H. Neumann and R.J. Zieve}
\affiliation{Physics Department, University of California at Davis}
\begin{abstract}
We present measurements of how a single vortex line in superfluid helium
interacts with a macroscopic bump on the chamber wall.  At a general level
our measurements confirm computational work on vortex pinning by a
hemispherical bump, but not all the details agree.  Rather than observing a
unique pin location, we find that a given applied velocity field can support
pinning at multiple sites along the bump, both near its apex and near its
edge. We also find that pinning is less favorable than expected.  A vortex
can pass near or even traverse the bump itself with or without pinning,
depending on its path of approach to the bump.
\end{abstract}
\maketitle

Vortex methods have appeared for decades in computational work on
classical fluids in both two and three dimensions \cite{Leonard,
Ojima10}.  They track the vorticity, which is the curl of the
velocity field, rather than the velocity itself.  The velocity can then be
extracted using the Biot-Savart law.  Vortex methods apply naturally to
situations where the vorticity is concentrated in particular regions
of the fluid; applications range from simulating trailing vortices of
aircraft \cite{Leonard} to doing time updates of the computer graphics
in video games \cite{Angelidis}.  A main benefit is that evaluating
the velocity field does not require tracking a fine grid of points.

A complication for vortex methods is the treatment of the vortex cores,
which in classical fluids change shape and size as vortex lines bend and
move.  Using a fixed size for the vortex cores, while more straightforward
in conception and implementation, is valid only when the core size is small
compared to all other length scales, including the local radius of curvature
of the vortex and the spacing between the computational points along the
vortex \cite{Fernandez}. This restriction does apply naturally in superfluid
${}^4$He, where the experimentally measured core radius is about 1.3\AA, 
small enough to be ignored on typical computational and experimental length
scales.  Conveniently, direct comparison between superfluid hydrodynamics
calculations and experiment becomes possible.  In several cases such
simulations accurately describe non-trivial experimental behaviors
\cite{Schwarzheli, smooth}.

Here we compare vortex pinning in experiment and calculation, finding
discrepancies that may indicate a need to modify  the computational
treatment of surfaces.  Our measurements track a single vortex in superfluid
helium interacting with a macroscopic bump. We compare to the computational
work of Schwarz \cite{Schwarz85}, for a hemispherical bump on an otherwise
flat wall.  Schwarz uses a flow field that far from the bump is uniform and
parallel to the wall.  He finds that if the vortex is swept into the
vicinity of the bump, the bump can capture and pin the vortex
\cite{Schwarz85}. If the flow velocity is large the vortex continues to
move, but for sufficiently low velocities it remains at the bump.  The
calculation uses no explicit pinning forces; rather, the stationary
configuration comes about entirely from the vortex settling into an
arrangement where the net velocity vanishes along its core.  For a given
flow velocity, the pinned vortex terminates at a unique position on the
bump, in the plane perpendicular to the flow. As the velocity increases, the
pin site moves out along the bump towards the wall.

Our experimental work confirms the general picture of a vortex pinning at a
bump, but we find some key differences.  First, we observe pinning near both
the apex and the edge of a bump, for the same applied velocity field. 
Second, in Schwarz's work even a vortex on a path that avoids the bump
can be pulled off course, encounter the bump, and pin to it.  By contrast,
we find that vortices only pin on the bump if they encounter it directly
and sufficiently close to its center.

\begin{figure}[b]
\begin{center}
\scalebox{.4}{\includegraphics{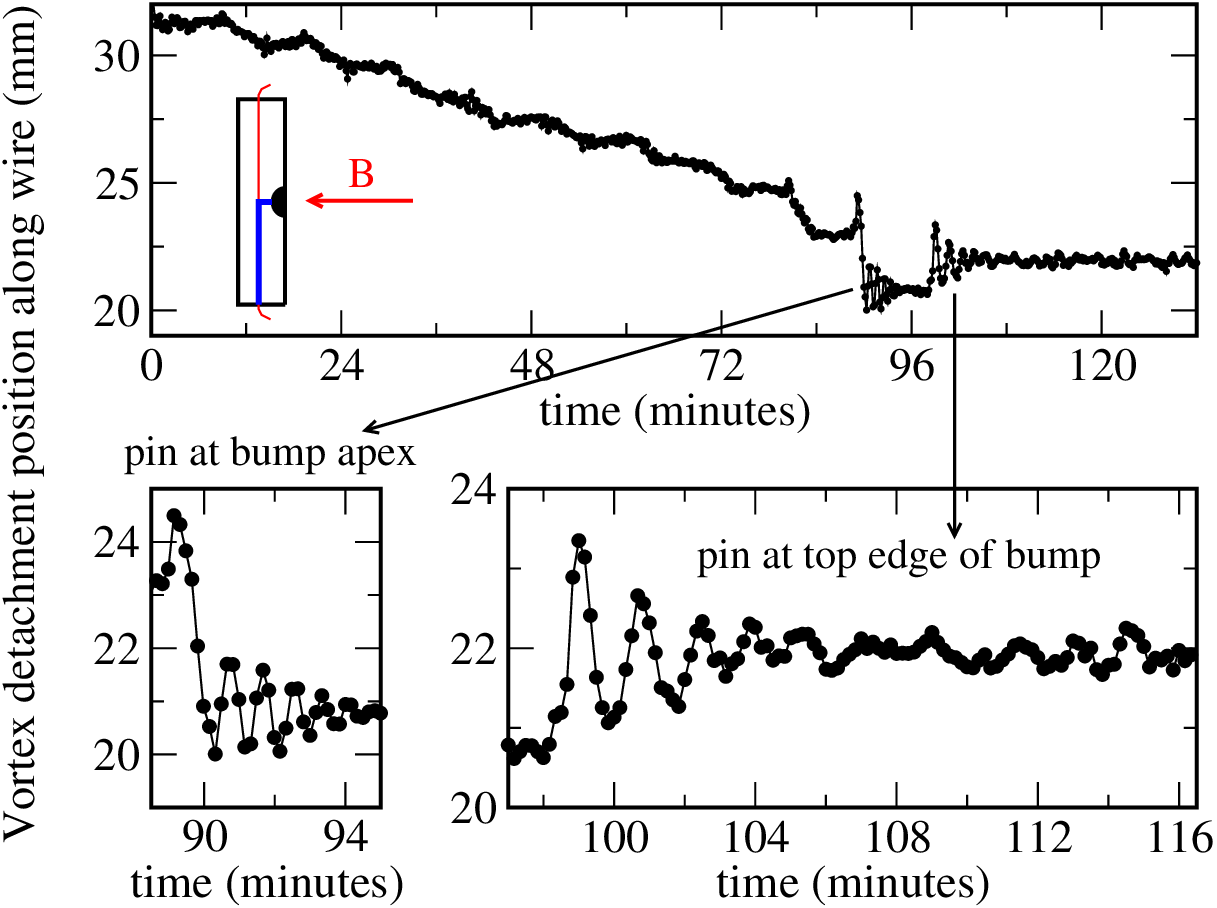}}
\caption{\small Top frame, inset: cell geometry,
with wire stretched near cell axis, a perpendicular magnetic field used to
excite and detect the wire's vibration, and a bump on the side wall. As drawn
a vortex extends along the wire from the bottom of the cell to the center,
then leaves the wire and terminates on the bump.  Top frame, main: motion of
the end of the vortex along the wire.  Initial oscillations correspond
to the free vortex precessing through the cell.  Pinning begins near 90
minutes.  Lower frames: expanded views of the Kelvin waves that occur while
the vortex is pinned.  The two frames have the same horizontal scale.
}
\label{f:pinrepin}
\end{center}
\end{figure}

Our data come primarily from a cylindrical cell of diameter 5.79 mm, with a
large bump midway along its length on the curved wall.  At its widest point
this bump has a roughly circular cross-section of diameter 3.05 mm, and its
apex protrudes a distance 1.27 mm from the circular wall.  As described
elsewhere \cite{smooth}, the cell is mounted on a pumped ${}^3$He cryostat
and filled with ${}^4$He through a small inlet hole in one end.  Our
measurements use a fine wire stretched vertically through the container, and
a constant horizontal magnetic field of order 25 mT.  The inset of Figure
\ref{f:pinrepin}a depicts this geometry. We pass a brief current pulse
through the wire; because of the static magnetic field, this creates a force
displacing the wire from its equilibrium position.  After the pulse ends,
the wire's tension causes it to vibrate and ultimately to settle back to its
equilibrium position.  As the wire moves, we monitor the emf induced across
it due to the horizontal magnetic field. 

We create vortices by rotating the cryostat at low temperatures, but we make
all our measurements with the cryostat stationary. Vorticity trapped around
all or part of the wire alters the observed vibration frequencies.  We focus
on the frequency splitting between the two lowest modes.  The earliest
measurements with a straight vibrating wire \cite{Vinen61} confirmed that
circulation is quantized in superfluid helium, since the frequency splitting
expected from a single quantum of circulation was strikingly stable. 
However, intermediate values of the frequency splitting also occur.  These
levels appear when a quantized vortex covers only a fraction of the wire and
hence has a reduced effect on the vibration frequencies.  From the observed
signal we can identify the spot where the vortex leaves the wire.  The
vortex must then continue through the fluid as a free vortex, and that free
portion moves at the local fluid velocity.  Since this motion often leads to
small adjustments to the position where the vortex leaves the wire, our
vibrating wire measurements allow us to track the free vortex portion.

In our physical experiment, the driving field is the flow field of the
trapped vertical vortex, which sweeps the free vortex segment around the
cell.  The free vortex terminates on the cell wall, and if it encounters the
bump during its precession, it can pin to the bump.  The resulting
experimental signature is that the circulation along the wire stabilizes;
the steady energy loss that we observe during precession ceases, as does the
oscillatory signal which corresponds to the circuit of the vortex around the
wire.

\begin{figure}[b]
\begin{center}
\scalebox{.42}{\includegraphics{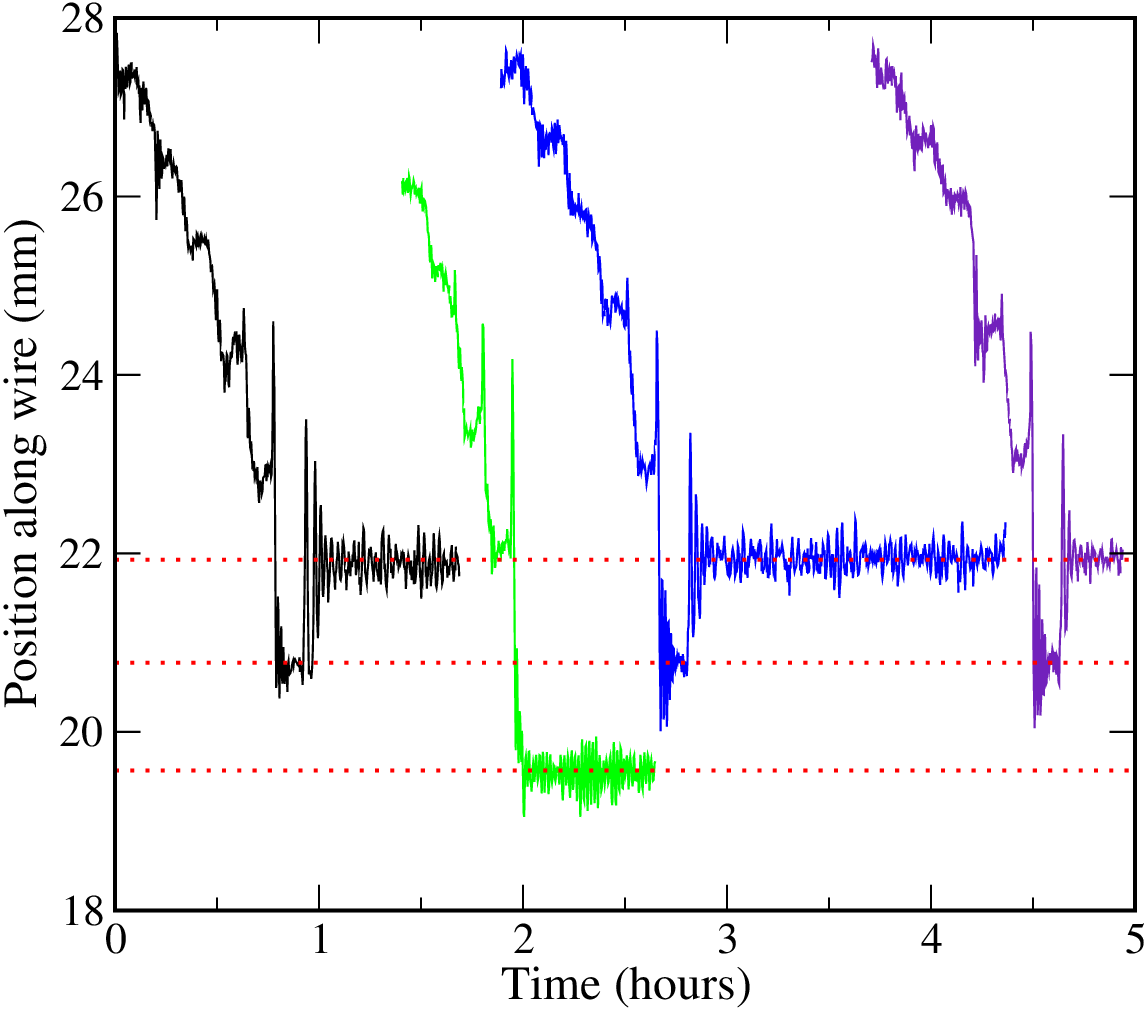}}
\caption{\small Several approaches of a vortex to the
bump, showing the reproducibility of the pin levels.  Each curve has
an arbitrary horizontal shift.  
}
\label{f:spikes}
\end{center}
\end{figure}

As a first indication that there is more than one metastable configuration
for the vortex on the bump, we observe not one but {\em three} closely
spaced levels for stable circulation.  Two appear in Figure
\ref{f:pinrepin}.  On approaching the bump, the vortex first pins with 20.7
mm of the wire's length covered by circulation.  After several minutes, this
level changes to 22.0 mm.  Figure \ref{f:spikes} includes the remaining
level, at 19.6 mm. The differences between these heights are comparable to
the widest radius of the bump.  Thus the middle level may correspond to
pinning at the apex of the bump, while the other two levels indicate pinning
close to the top and bottom of the circle where the bump meets the wall.

Kelvin waves during the pins confirm that the three levels correspond to
pinning at different parts of the bump.  The Kelvin waves appear as rapid
oscillations superimposed on the steady circulation. We have shown
previously that Kelvin waves can be excited by our vibrating wire itself,
particularly when the other end of the vortex is fixed \cite{energyloss}. 
From our earlier work, the observed oscillation frequencies correspond to
the longest-wavelength modes with the vortex fixed at the cell wall and free
to move vertically along the wire.  (Empirically, since we observe
oscillations at the Kelvin wave frequency, the vortex must not be fixed at
the wire.) The wire used for the present measurements exhibits Kelvin waves
often, and their frequencies provide key geometric information.

Frames b) and c) of Figure \ref{f:pinrepin} expand the oscillations visible
at each of the two stable circulation levels.  The horizontal axis has the
same scale in both cases, illustrating clearly that the oscillations at the
first pin have a higher frequency.  The periods are about 52 seconds for
the first pinned level and 101 seconds for the second level.  The period
repeats to within the uncertainty of about two seconds for all pinning
events at the same level.

The ideal Kelvin wave period in the long-wavelength limit is $$T =
\frac{2\lambda^2}{\kappa \ln\frac{\lambda}{2a_o}},$$ where $a_o=1.3\times
10^{-7}$ mm is the vortex core radius, $\kappa=9.97\times 10^{-2}$
mm$^2$/s is the circulation of the vortex, and $\lambda$ is the wavelength
of the Kelvin mode \cite{Donnelly}. For a vortex with one end pinned, the
lowest-frequency mode has wavelength four times the length of the free
vortex segment.  If the pin site is at the bump apex, then the free
vortex length is 1.6 mm and the corresponding period is $T=48$ seconds,
very close to the observed value at the middle pin level.  A pin site
at the edge of the bump, with vortex length equal to the cell radius
of 2.9 mm, gives a period of 164 seconds.  Here the agreement is not
especially good, although that could mean simply that the vortex does not
pin precisely at the edge of the bump.  Indeed, if we model the exposed
half of the bump as half of an ellipsoid, then the straight-line path
from the center of the cell to the bump edge passes {\em through} the
bump.  The observed period of 101 s suggests a vortex length of 2.3 mm.
For our bump, this would occur at a distance 1.2 mm from the bump axis,
which is still quite close to the edge laterally.

The Kelvin waves at the third pin level have an intermediate period
near 62 seconds.  This corresponds to a vortex length of 1.76 mm, which
would place the pin site 0.97 mm off the axis.  Since the glue ran down
in this direction when the bump was affixed to the cell wall, this edge
does have a more gradual height change and the observed values for the
pin location and Kelvin wave period are plausible.

Figure \ref{f:spikes} is a compendium of several vortex pinning events.
In three cases a vortex pins to the bump apex, then works free
spontaneously after about 8 minutes and shifts to a site closer to the
bump edge.  By contrast, the pins at both the top and bottom edges of the
bump never come free without deliberate heating on our part to dislodge
the vortex.  Thus it appears that for this geometry pinning to the bump
apex is less stable than pinning closer to the bump edge.  This temporary
pinning does not occur in the computational work \cite{Schwarz85}.
One possible reason for the difference is the much lower dissipation in
our actual experiment. The computational vortices are far more able to
dispose of excess length through Kelvin oscillations and the resulting
energy loss; in our experiment, the oscillations may continue at high
amplitude and eventually dislodge the vortex.

As noted above, both the pin levels and the Kelvin wave periods for each
level are highly reproducible, strongly suggesting that vortices repeatedly
pin at the same few spots. However, the initial approaches of the vortices
are far from identical.  Each trace begins with a few cycles of a slow
oscillation, with period about 10 minutes.  This feature corresponds to
precession about the wire of the free vortex segment.  Minima and maxima in
this precession indicate particular directions of the free vortex within the
cell.  Since the heights of the minima and maxima vary from one trace to
another, the different vortices must approach the bump along different
trajectories but still reach the same pin locations.  This cannot occur
purely from microscopic roughness at the pin site; macroscopic energy
considerations must play a role in guiding the vortex to the site.
Similarly, the three vortices that move from a pin site at the bump apex to
one at the bump edge trace out different paths. For example, as Figure
\ref{f:spikes} shows, the number of large-amplitude Kelvin wave periods
before the vortex settles is different for each trace.

One unusual energy consideration that could contribute to the existence
of multiple pin sites is the Gaussian curvature of the surface
\cite{VitelliNelson, VitelliTurner}.  Regions of negative Gaussian
curvature are predicted to be more favorable for defects than regions of
positive curvature.  The arguments rely on energetics of two-dimensional
systems, and a proposed test in superfluid helium involves a thin layer
of superfluid \cite{VitelliTurner}.  In a three-dimensional system,
surface energy terms are likely to be much smaller than bulk terms, but
they may still provide an incremental contribution that leads to a metastable pin location.

A Gaussian curvature effect could explain the particularly unexpected location of one of our pin sites.  Since the vortex must
be perpendicular to the bump at its pin location, any pin site other
than the bump apex requires the vortex to curve.  For a stable pinned
vortex, the velocity field produced by this curvature exactly cancels
the applied velocity field.  Following this logic, in our experiment we
expect the stable pin site to lie towards the bottom edge of the bump.
Yet our measurements repeatedly show a vortex pinning near the
{\em top} edge, where the self-induced velocity near
the bump augments the applied velocity.  A pinning force, perhaps
deriving from the energetics of Gaussian curvature, is needed to retain
the vortex at this spot.

Another result from Schwarz's calculations is that if the fluid velocity is
low enough for a vortex to pin, then the vortex will do so as long as it
moves along a path that passes within about one bump radius of the edge of
the bump \cite{Schwarz85}.  The distortion of the velocity field by the bump
pulls the vortex inward until it encounters the bump. By contrast, we find
that whether or not the vortex pins depends strongly on its exact approach
to the bump.  The spikes in Figure \ref{f:spikes} in the two or three
precession cycles immediately before a vortex pins indicate that the vortex
is moving over the surface of the bump but not pinning. If the end of the
vortex traverses the bump, then the length of the free vortex shrinks
abruptly.  The length of circulation trapped on the wire increases
sharply to compensate, causing the spike in our data.  As the level at
which the vortex encounters the bump approaches the bump center, the spike
magnitude increases.  For certain approach paths the vortex pins at the
bump, although for others with essentially the same applied velocity field
the vortex does not pin. The difference from the simulations may stem from
our lower dissipation or from our applied velocity field's not being uniform.

\begin{figure}[b]
\begin{center}
\scalebox{.42}{\includegraphics{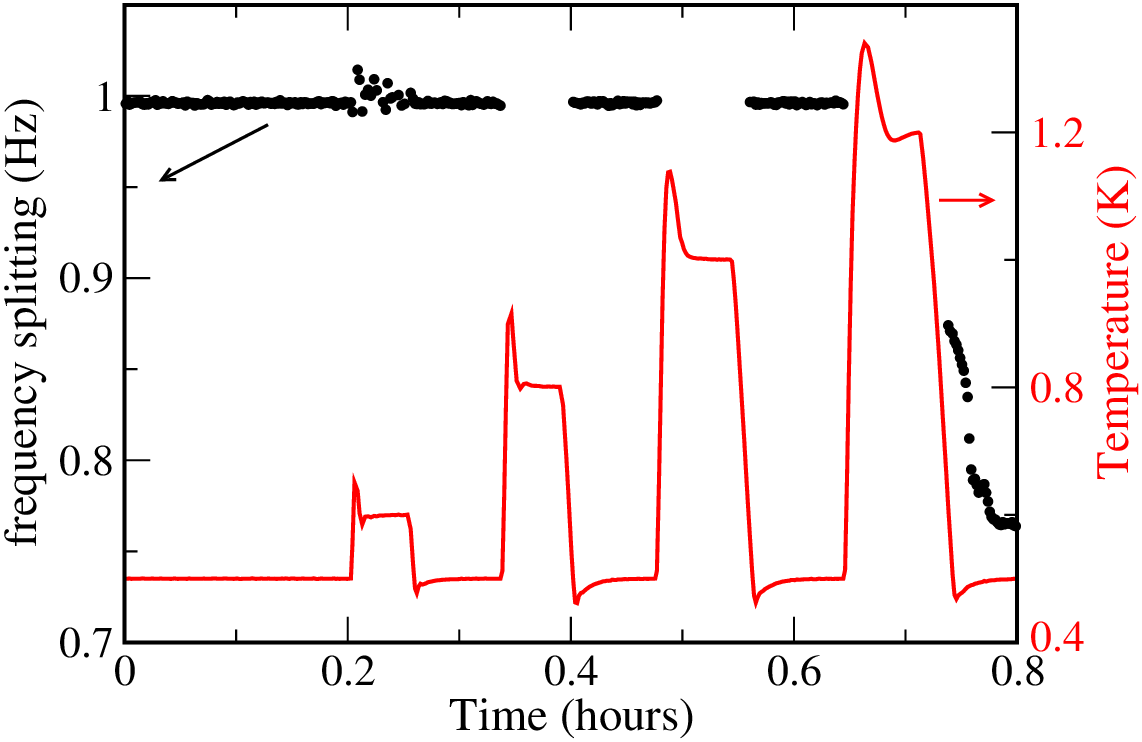}}
\caption{\small Thermal cycling to depin a vortex from the
wire; see text for details.  Right (red) axis is temperature; left
(black) axis is the frequency splitting of the lowest modes of the
wire.  The trace begins with a single quantum of circulation
surrounding the entire wire; coincidentally, the frequency splitting
for this circulation happens to be very close to 1 Hz.
}
\label{f:heat}
\end{center}
\end{figure}

A different geometry provides additional evidence that a
vortex can pin at the edge or apex of a bump.  In this cell the bump is
attached not to the side wall but to one of the endcaps, with the wire
glued to the bump apex.  Thus a vortex that is trapped along the entire
length of the wire terminates at the apex.  Alternatively, a vortex that
leaves the wire shortly before it reaches the bump and traverses the
superfluid as a free vortex can pin near the edge of the bump. We cannot
distinguish these two situations directly.  In principle the beat
frequency is slightly smaller if the vortex does not cover the entire
wire.  Unfortunately, for any reasonable configuration the missing length
would be quite small, and the frequency difference would be unobservable. 
In part this is because the measurement sensitivity is vastly reduced near
the ends of the wire compared to the middle.  The wire's displacement
during vibration has nodes at the ends of the cell, so circulation near
the ends has little influence on the wire's motion.

However, we do observe an indirect signature related to the stability of
pinned vortices.  Figure \ref{f:heat} illustrates how we test stability.
We provide thermal energy by raising the temperature of the cryostat.
At the higher temperatures the damping of the vibrating wire is too high
to extract the circulation, so after a few minutes we cool the cryostat
and check whether the circulation level has changed.  If it has not,
we heat again to a slightly higher temperature, repeating until the
vortex depins.  After the vortex dislodges, the ensuing precession signal
indicates that the vortex now lies along only part of the wire's length,
with one end terminating on the cylindrical wall.

\begin{figure}[t]
\begin{center}
\scalebox{.42}{\includegraphics{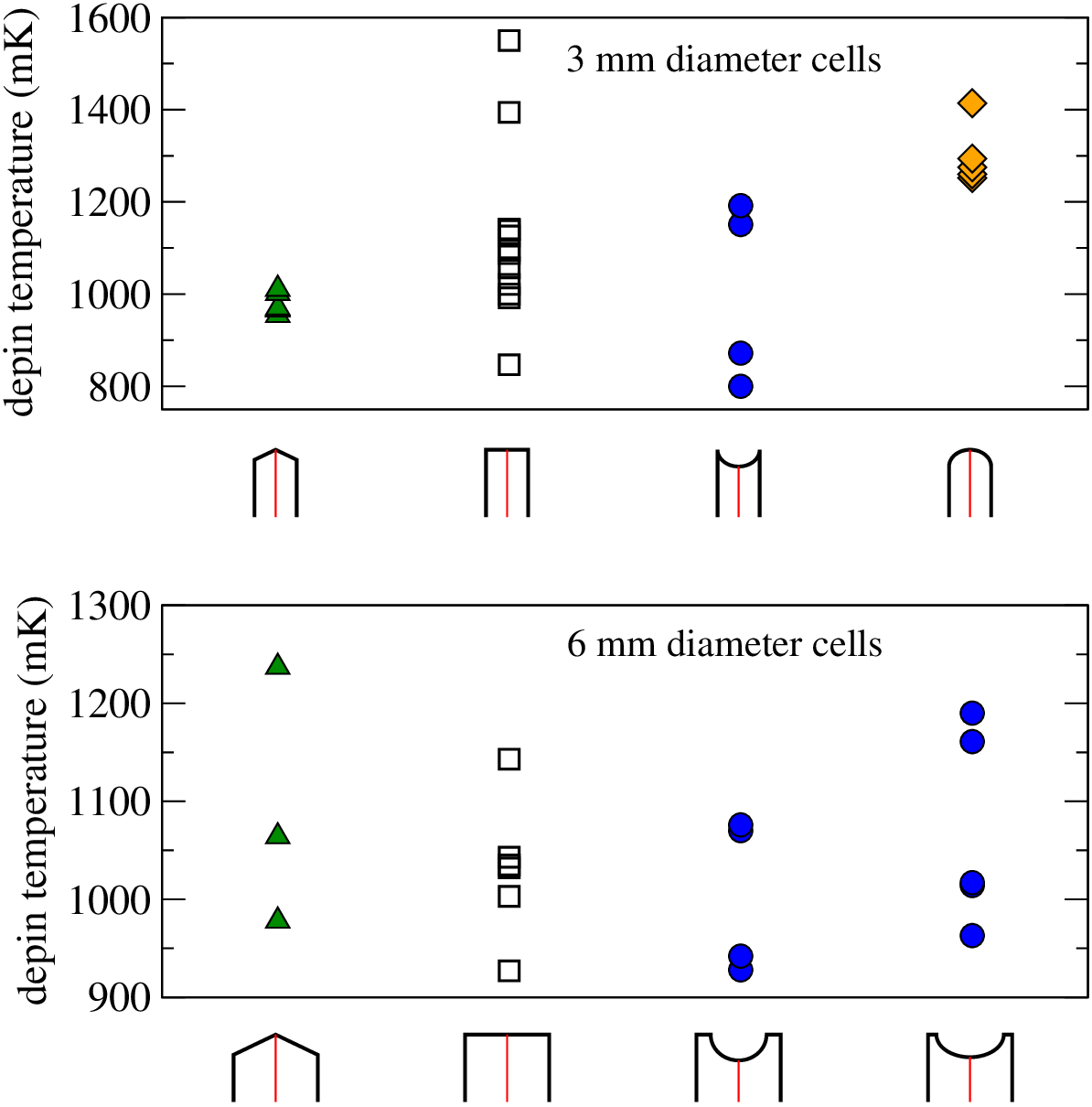}}
\caption{\small Depin temperatures for eight cells.  The
shape of the cell's end cap (cone, flat, bump, or hemispherical
indentation) is sketched below the data points for that cell. All caps
have cylindrical symmetry. Each point corresponds to a single pinned
vortex and indicates the temperature reached when the vortex left the
wire.  For the vortex shown in Figure \ref{f:heat}, this temperature is
1.34 K, the maximum temperature of the final heating cycle. }
\label{f:depinT}
\end{center}
\end{figure}

Figure \ref{f:depinT} shows the temperatures at which vortices depin
for several cells with different endcap configurations.  Each point
represents a single pinned vortex, with the temperature derived from a
heating sequence such as that in Figure \ref{f:heat}.  Many of the
cells show a large spread of depinning temperatures for different
vortices, possibly depending on details of the heating cycles or on the
interplay between the circulation and the end of the wire. Nonetheless,
some patterns emerge, such as the high stability of vortices in the cell
that ends with the hemispherical indentation. Three of the eight cells have
endcaps with bumps.  In those three, but not in any others, the depin
temperatures cluster into two groupsm which may correspond to pinning at the
bump apex and at the bump edge.  A vortex pinned at the bump edge is already
close to the outer wall of the cylinder, and we expect it to have a smaller
energy barrier to overcome in depinning compared to a vortex that follows
the entire wire. Hence the lower-temperature depins for each bump could
occur for vortices pinning near the bump edge, while those at higher
temperature signify vortices pinning at the apex. We note that the thin cell
with a bump has some particularly low depin temperatures and also has the
closest approach of the bump edge to the outer wall of the cell.

Our measurements agree with the qualitative picture of vortex pinning
that arises from numerical simulations.  However, given the past successes
of superfluid hydrodynamics calculations, the discrepancies may indicate
additional physics not accounted for in the computations.  We observe
multiple metastable pin sites on a single convex bump.  In addition, vortices
passing near the bump do not spiral inward and pin.  On the contrary,
even vortices that encounter the bump directly sometimes pass over it
without pinning.  We are pursuing further computational and experimental
work to resolve these issues.

We acknowledge funding from UC Davis.

\end{document}